\documentclass[abstract=on,notitlepage,superscriptaddress,nofootinbib,aps,preprint,prd]{revtex4}
\usepackage{srcltx}
\usepackage{epsf}
\usepackage{amsthm}
\usepackage{amsfonts}
\usepackage{amsmath}
\usepackage{mathrsfs}
\usepackage{epsfig}
\usepackage{enumerate}
\usepackage{lipsum}
\usepackage{graphicx}
\usepackage{epsf}
\usepackage{subfig}
\usepackage{color}
\usepackage{caption}
\usepackage{subfig}
\usepackage[dvipsnames]{xcolor}
\usepackage{epstopdf}
\usepackage{float}
\usepackage{dcolumn}
\usepackage{hyperref}
\usepackage{amsthm}
\usepackage{graphicx, amsmath, amssymb, bm}

\newcommand{\be}{\begin{equation}}
\newcommand{\ee}{\end{equation}} 
\newcommand{\eei}{\end{equation}\indent\indent}
\newcommand{\bc}{\begin{center}}
\newcommand{\ec}{\end{center}}
\newcommand{\ber}{\begin{eqnarray*}}
\newcommand{\ear}{\end{eqnarray*}}
\newcommand{\ba}{\begin{array}}
\newcommand{\ea}{\end{array}}
\def\n{\nonumber}

\newcommand{\sss}[1][0.035cm]{\hspace*{#1}}

\newcommand{\bea}{\begin{eqnarray}}
\newcommand{\eea}{\end{eqnarray}}

\newcommand{\ei}{\end{itemize}}

\begin{document}


\title{A semi-tetrad decomposition of the Kerr spacetime}



\author{Chevarra Hansraj}\email[]{chevarrahansraj@gmail.com}
\affiliation{Astrophysics and Cosmology Research Unit, School of Mathematics, Statistics and Computer Science,
	University of KwaZulu--Natal, Private Bag X54001, Durban 4000, South Africa}
\affiliation{DST-NRF Centre of Excellence in Mathematical and Statistical Sciences (CoE-MaSS)}
\author{Rituparno Goswami}\email[]{Goswami@ukzn.ac.za}
\affiliation{Astrophysics and Cosmology Research Unit, School of Mathematics, Statistics and Computer Science,
	University of KwaZulu--Natal, Private Bag X54001, Durban 4000, South Africa}
\author{Sunil D. Maharaj}\email[]{maharaj@ukzn.ac.za}
\affiliation{Astrophysics and Cosmology Research Unit, School of Mathematics, Statistics and Computer Science,
	University of KwaZulu--Natal, Private Bag X54001, Durban 4000, South Africa}


\begin{abstract}
	In this paper we perform a semi-tetrad decomposition of the Kerr spacetime. We apply the 1+1+2 covariant method to the Kerr spacetime in order to describe its geometry outside the ergoregion. As a result we are able to explicitly write down the 1+1+2 Kerr quantities, and the evolution and propagation equations they satisfy. This formalism allows us to present the kinematic and dynamic quantities in a transparent geometrical manner; and also to highlight the role of vorticity. To our knowledge, using the 1+1+2 formalism to investigate the Kerr spacetime is a novel approach and this provides new insights into the spacetime geometry in an easier manner than alternate approaches. Furthermore we make corrections to earlier equations in the 1+1+2 formalism applied to the Kerr spacetime.
	
\end{abstract}


\maketitle


\section{Introduction}
The Kerr spacetime, discovered in 1963 by Roy Kerr, is extremely relevant to the understanding of black hole physics and modern astrophysics. To a very good approximation, the spacetime near each rotating black hole in the observable universe is given by a unique exact solution of Einstein's vacuum field equations which is the Kerr metric \cite{kerr}. The Kerr metric extends the Schwarzschild and the Reissner-Nordstrom metrics to include angular momentum. In this way rotation can also be included in the model. It is crucially important to study the effects of gravity in the Kerr geometry. Then only will it be possible to build a model of a rotating isolated body in general relativity which is an unsolved problem in astrophysics.

Examples of well known tetrad or semi-tetrad methods are the complex null tetrads of Newman and Penrose, the 1+3 covariant approach developed by Ehlers \cite{ehlers} and Ellis \cite{ellis1} and the 1+1+2 covariant approach developed by Clarkson and Barrett \cite{clarksonbarrett}. The 1+3 formalism has generated new results in areas like
gauge-invariant study (\cite{goswami}, \cite{ellis2}), the cosmic microwave background \cite{challinor} and specific spacetimes (\cite{ellis3}-\cite{bianchi}). An extension of the 1+3 covariant approach is the 1+1+2 covariant approach which has generated new results in locally rotationally symmetric spacetimes in general relativity \cite{singh} and $f(R)$ gravity \cite{nzioki}, and spacetimes with conformal symmetry \cite{hansraj}. 

Through the years particular interest in the Kerr geometry has been generated through various studies (\cite{chandrasekhar}-\cite{kalnins}). In \cite{chandrasekhar} the Newman-Penrose formalism involving null tetrads was used to explore the solution of Maxwell's equations in the Kerr geometry. In this paper we use both the 1+3 and 1+1+2 covariant approaches to describe the Kerr spacetime geometry. The advantage of using these approaches is that the physics and geometry of the spacetime are described by tensor quantities and relations which remain valid in all coordinate systems. The purpose of using these methods is to extract the geometrical features of the spacetime in an easier manner as the geometric variables have well defined physical interpretations. These geometric variables have been defined explicitly in this paper. A partial study of the Kerr metric in the 1+3 formalism was done in \cite{frolovnovikov}; we complete this analysis and write down the full set of 1+3 equations for the Kerr metric. Noteworthily we apply the 1+1+2 formalism to the Kerr spacetime and explicitly write down the 1+1+2 Kerr geometrical quantities and the evolution and propagation equations they satisfy. To our knowledge using the 1+1+2 formalism to investigate the Kerr spacetime is a novel approach and it provides new insights into the spacetime geometry. All quantities and equations have been validated with the mathematical software Maple and GRTensor.

The paper is structured as follows: In the following section we define the Kerr metric and consider its key features. In section 3 and Appendix A we briefly review the 1+3 formalism. Then in section 4 we explicitly write down the 1+3 Kerr quantities and the equations they satisfy. Section 5 and Appendix B contain a brief review of the 1+1+2 formalism. In section 6 and Appendix C we apply the 1+1+2 formalism to the Kerr spacetime, a novel approach, and explicitly write down the 1+1+2 Kerr geometrical quantities and the evolution and propagation equations. Some errors in the 1+1+2 formalism equations found earlier are identified and corrected for the Kerr spacetime. Note that these corrections do not alter the linear perturbation results presented in \cite{clarksonbarrett} and \cite{clarkson2007}. Concluding remarks are made in section 7.
 
\section{The Kerr metric}
The Kerr metric describes the vacuum, stationary axisymmetric solution corresponding to stationary rotating black holes and depends on angular momentum ${(a)}$ and mass ${(m)}$ as parameters.
One of the key features of the Kerr spacetime geometry is that it is Ricci flat $\left(R_{ab} = 0\right)$. Additionally, there are three off-diagonal terms in the very first version of the line element \cite{kerr}. Considering the Kerr spacetime in Boyer-Lindquist coordinates \cite{boyerlindquist} which involves a coordinate substitution, results in only one off-diagonal term. Starting from the Boyer-Lindquist coordinates $\left(t, r, \theta, \varphi\right)$, a new coordinate ${\chi = \cos\theta}$ so that $\chi \in \left[-1 ; 1\right]$ is introduced, and we can write the Kerr spacetime in terms of ``rational polynomial" coordinates $\left(t, r, \chi, \varphi\right)$ as
\begin{eqnarray}
ds^{2} &=& \left(\frac{2mr}{r^{2} + a^{2}\chi^{2}} - 1\right)dt^{2} - \left[\frac{4mar\left(1 - \chi^{2}\right)}{r^2 + a^{2}\chi^{2}}\right]dt d\varphi \nonumber\\
& & + \left(\frac{r^{2} + a^{2}\chi^{2}}{r^2 - 2mr + a^{2}}\right)dr^{2} + \left(\frac{r^{2} + a^{2}\chi^{2}}{1-\chi^{2}}\right)d\chi^{2} \nonumber\\
& & +\left(1 - \chi^{2}\right)\left[r^{2} + a^{2} + \frac{2ma^{2}r\left(1 - \chi^{2}\right)}{r^{2} + a^{2}\chi^{2}}\right] d\varphi^{2}.
\end{eqnarray}
For the purpose of this paper we will consider the Kerr spacetime in rational polynomial coordinates. The advantage of using these coordinates is that it eliminates trigonometric functions so that computational calculations can be performed more efficiently. We note that we are studying the Kerr metric outside the ergoregion. This is because our description of the unit vector ${u^{a}}$ \eqref{defU} entirely depends on $r^{2} - 2mr + a^{2}\chi^{2}$ which is solved to work out the ``stationary limit" surfaces between which it is impossible to stand still. Recent studies on the Kerr spacetime have been done in the context of unit-lapse forms \cite{baines}, vortex forms \cite{visser}, its topology \cite{shatskiy} and coding simulations \cite{bambi}. For a comprehensive review of the Kerr spacetime the reader is referred to \cite{visserbook}.


\section{1+3 formalism}

In the 1+3 formalism, the timelike unit vector ${u^{a}}$ ${\left(u^{a} u_{a} = -1\right)}$ is split in the form ${\mathcal{R} \otimes \mathcal{V}}$, where ${\mathcal{R}}$ is the timeline along ${u^{a}}$ and ${\mathcal{V}}$ is the 3-space perpendicular to ${u^{a}}$. The 1+3 covariantly decomposed spacetime is represented by
\begin{equation}
 g_{ab} = h_{ab} - u_{a} u_{b},
\end{equation}
where ${h_{ab}}$ is a tensor that projects onto the rest space of an observer moving with 4-velocity ${u^{a}}$. The covariant time derivative along the observers' worldlines, denoted by `\,${^{\cdot}}$\,', is defined using the vector ${u^{a}}$, as
\begin{equation} 
\dot{Z}^{a ... b}{}_{c ... d} = u^{e}\nabla_{e} Z^{a ... b}{}_{c ... d},
\end{equation} 
for any tensor ${Z^{a...b}{}_{c...d}}$. The fully orthogonally projected covariant spatial derivative, denoted by `\,${D}$\,', is defined using the spatial projection tensor ${h_{ab}}$, as
\begin{equation}
D_{e} Z^{a...b}{}_{c...d} = h^r{}_{e} h^p{}_{c}... h^q{}_{d} h^a{}_{f}... h^b{}_{g}\nabla_{r} Z^{f...g}{}_{p...q},
\end{equation}
with total projection on all the free indices. The covariant derivative of the 4-velocity vector ${u^{a}}$ is decomposed irreducibly as follows
\begin{eqnarray}
\nabla_{a} u_{b} &=& -u_{a} \dot{u}_{b} + \frac{1}{3}h_{ab}\Theta + \sigma_{ab} + \varepsilon_{abc} \omega^{c},
\end{eqnarray}
where ${\dot{u}_{b}}$ is the acceleration, ${\Theta}$ is the expansion of ${u_{a}}$, ${\sigma_{ab}}$ is the shear tensor, ${\omega^{a}}$ is the vorticity vector representing rotation and ${\varepsilon_{abc}}$ is the effective volume element in the rest space of the comoving observer. A more detailed review of the formalism can be found in \cite{ellis5}.

\section{The 1+3 Kerr quantities and equations}
We apply the 1+3 formalism to the Kerr metric and find the following quantities which have been confirmed by the mathematical software Maple and GRTensor.
The timelike unit vector is defined as 
\begin{eqnarray} \label{defU}
u^{a} &=& \left[\sqrt{\frac{r^{2} + a^{2}\chi^{2}}{r^{2} - 2mr + a^{2}\chi^{2}}}, 0, 0, 0\right]  \text{where } u^{a}u_{a} = -1.
\end{eqnarray}
We re-emphasise that this definition of ${u^{a}}$ is applicable outside the ergoregion. Inside the ergoregion the formalism can be applied but with a different choice of ${u^{a}}$.

The set of the Kerr 1+3 geometric variables is given by 
\begin{eqnarray}
&&\left\{\Theta, \sigma_{ab}, \omega^{a}, \omega_{ab}, \dot{u}_{a}, E_{ab}, H_{ab}\right\},
\end{eqnarray}
and they have the following values
{\allowdisplaybreaks{
\begin{eqnarray}
\Theta &=& D^{a}u_{a} = 0, \\
\sigma_{ab} &=& D_{(a} u_{b)} - \frac{1}{3}\Theta h_{ab} = 0, \\
\omega_{ab} &=& D_{[a}u_{b]} \nonumber \\
					  &=& \begin{bmatrix}
0 & 0 & 0 & 0\\
0 & 0 & 0 & \frac{ma\mathcal{M}\mathcal{D}}{\sqrt{\mathcal{J}^3\mathcal{K}} }\\
0 & 0 & 0 & \frac{2mar\chi\mathcal{L}}{ \sqrt{\mathcal{J}^3\mathcal{K}} }\\
0 & -\frac{ma\mathcal{M}\mathcal{D}}{\sqrt{\mathcal{J}^3\mathcal{K}} } & -\frac{2mar\chi\mathcal{L}}{\sqrt{\mathcal{J}^3\mathcal{K}}} & 0
\end{bmatrix},\\
\omega^{a} &=& \frac{1}{2}\varepsilon^{abc}\omega_{bc} \nonumber\\
		 &=& \left[0, -\frac{2mar\chi\mathcal{L}}{\mathcal{J}\mathcal{K}^2}, \frac{ma\mathcal{M}\mathcal{D}}{\mathcal{J}\mathcal{K}^2},0 \right], \\		 
 \dot{u}^{a} &=& u^{b}\nabla_{b}u^{a} \nonumber\\
				&=&  \left[0, -\frac{m\mathcal{D}\mathcal{L}}{\mathcal{J}\mathcal{K}^2}, -\frac{2ma^{2}r\chi\mathcal{M}}{\mathcal{J}\mathcal{K}^2},0 \right], \\	 
E_{ab} &=& C_{acbd}u^{c}u^{d} \nonumber\\
&=& \begin{bmatrix}
0 & 0 & 0 & 0\\
0 & -\frac{mr\mathcal{P}\left(a^{2}\chi^{2} - 3a^{2} + 4mr - 2r^{2}\right)}{\mathcal{J}\mathcal{K}^2\mathcal{L}} & \frac{3ma^{2}\chi \mathcal{Q}}{\mathcal{J}\mathcal{K}^2} & 0\\
0 & \frac{3ma^{2}\chi \mathcal{Q}}{\mathcal{J}\mathcal{K}^2} & - \frac{mr\mathcal{P}\left(2a^{2}\chi^{2} - 3a^{2} -r^{2} + 2mr\right)}{\mathcal{J}\mathcal{K}^2\mathcal{M}} & 0 \\
0 & 0 & 0 & \frac{mr\mathcal{LP}\mathcal{M}}{\mathcal{J}\mathcal{K}^2} 
\end{bmatrix},\\
H_{ab} &=& \frac{1}{2}\eta_{abef}C^{ef}{}_{cd}u^{c}u^{d} \nonumber\\
&=& \begin{bmatrix}
0 & 0 & 0 & 0\\
0 & -\frac{ma\chi \mathcal{Q}\left(a^{2}\chi^{2} - 3a^{2} + 4mr -2r^{2}\right)}{\mathcal{J}\mathcal{K}^2\mathcal{L}} & -\frac{3mar\mathcal{P}}{\mathcal{J}\mathcal{K}^2} & 0\\
0 & -\frac{3mar\mathcal{P}}{\mathcal{J}\mathcal{K}^2} & -\frac{ma\chi \mathcal{Q}\left(2a^{2}\chi^{2} - 3a^{2} -r^{2} + 2mr\right)}{\mathcal{J}\mathcal{K}^2\mathcal{M}} & 0 \\
0 & 0 & 0 & \frac{ma\chi \mathcal{LQ}\mathcal{M}}{\mathcal{J}\mathcal{K}^2} 
\end{bmatrix},
\end{eqnarray}}\\
where 
\begin{eqnarray}
\mathcal{D} &=& -r^{2} + a^{2}\chi^{2},  \nonumber\\
\mathcal{J} &=& r^{2} - 2mr + a^{2}\chi^{2},  \nonumber\\ 
\mathcal{K} &=& r^{2} + a^{2}\chi^{2},  \nonumber\\
\mathcal{L} &=& r^{2} - 2mr + a^{2}, \nonumber\\
 \mathcal{M} &=& -1 + \chi^{2},  \nonumber\\
  \mathcal{P} &=& 3a^{2}\chi^{2} - r^{2},  \nonumber\\
  \mathcal{Q} &=& a^{2}\chi^{2} - 3r^{2}. \nonumber
\end{eqnarray}}
We note that the rate of volume expansion scalar ${(\Theta)}$ and the shear ${(\sigma_{ab})}$ for the chosen unit timelike vector are zero. A 1+3 decomposition on the Kerr spacetime was partially investigated in \cite{frolovnovikov}. We highlight that the quantities of acceleration ${\left(\dot{u}^{a}\right)}$ and vorticity ${\left(\omega_{ab}\right)}$ are consistent with the findings of Frolov and Novikov \cite{frolovnovikov}. 

Now we present the full set of 1+3 Kerr equations. These equations as well as the relevant identities in Appendix A have been checked in Maple. 
The first set of equations are derived from the Ricci identities. We write down three propagation equations and three constraint equations.

\subsection{Propagation equations I:}
\noindent Raychaudhuri equation
\begin{equation}
D_{a}\dot{u}^{a} = -\dot{u}_{a}\dot{u}^{a} - 2\omega_{a}\omega^{a}.
\end{equation}
Vorticity propagation equation
\begin{equation}
\dot{\omega}^{<a>} - \frac{1}{2}\varepsilon^{abc}D_{b}\dot{u}_{c} = 0.
\end{equation}
Shear propagation equation
\begin{equation}
D^{<a}\dot{u}^{b>} = -\dot{u}^{<a}\dot{u}^{b>} + \omega^{<a}\omega^{b>} + E^{ab}.
\end{equation}
\subsection{Constraint equations I:}
{}
\begin{eqnarray}
0 &=& \varepsilon^{abc}\left[D_{b}\omega_{c} + 2\dot{u}_{b}\omega_{c}\right]. \\
0 &=& D_{a}\omega^{a} - \dot{u}_{a}\omega^{a}. \\
0 &=& H^{ab} + 2\dot{u}^{<a}\omega^{b>} + D^{<a}\omega^{b>}.
\end{eqnarray}
The second set of equations arise from the once-contracted Bianchi identities. We write down two further propagation equations and two further constraint equations.
\subsection{Propagation equations II:}
${\dot{E}}$-equation
\begin{equation}
\dot{E}^{<ab>} - \varepsilon^{cd<a}D_{c}H^{b>}{}_{d} = \varepsilon^{cd<a}\left[2\dot{u}_{c}H^{b>}{}_{d} + \omega_{c}E^{b>}{}_{d}\right].
\end{equation}
${\dot{H}}$-equation
\begin{equation}
\dot{H}^{<ab>} + \varepsilon^{cd<a}D_{c}E^{b>}{}_{d} = -\varepsilon^{cd<a}\left[2\dot{u}_{c}E^{b>}{}_{d} - \omega_{c}H^{b>}{}_{d}\right].
\end{equation}
\subsection{Constraint equations II:}
\noindent
(div ${E}$)-equation
\begin{equation}
0 = (C_{4})^{a} = D_{b}E^{ab} - 3\omega_{b}H^{ab}. 
\end{equation}
 (div ${H}$)-equation
\begin{equation}
0 = (C_{5})^{a} = D_{b}H^{ab} + 3\omega_{b}E^{ab}.
\end{equation}


\section{1+1+2 formalism}

In the 1+1+2 formalism, the 3-space ${\mathcal{V}}$ is now further split by introducing the unit vector ${e^{a}}$ orthogonal to ${u^{a}}$ ${\left(e^{a} e_{a} = 1, u^{a} e_{a} = 0\right)}$. The 1+1+2 covariantly decomposed spacetime is given by 
\begin{equation}\label{2.Nab}
g_{ab} = -u_{a} u_{b} + e_{a} e_{b} + N_{ab},
\end{equation}
where ${N_{ab}}$ ${\left(e^{a} N_{ab} = 0 = u^{a} N_{ab}, \,N^{a}{}_{a} = 2\right)}$ projects vectors orthogonal to ${u^{a}}$ and ${e^{a}}$ onto 2-spaces called `sheets.' 
We introduce two new derivatives for any tensor ${\Psi_{a...b}{}^{c...d}}$:
\begin{eqnarray}
\label{hatderiv}
\hat{\Psi}_{a...b}{}^{c...d} &\equiv& e^{f} D_{f} \Psi_{a...b}{}^{c...d}, \\
\label{deltaderiv}
\delta_{f}\Psi_{a...b}{}^{c...d} &\equiv& N_{f}{}^{j} N_{a}{}^{l} ... N_{b}{}^{g} N_{h}{}^{c} ... N_{i}{}^{d}  D_{j}\Psi_{l...g}{}^{h...i}, 
\end{eqnarray}
defined by the congruence ${e^{a}}$. The hat-derivative (\ref{hatderiv}) is the spatial derivative along the ${e^{a}}$ vector field in the surfaces orthogonal to ${u^{a}}$ and the delta-derivative (\ref{deltaderiv}) is the projected spatial derivative onto the 2-sheet, with projection on every free index.

Taking ${e^{a}}$ to be arbitrary, the 1+3 kinematical quantities and anisotropic fluid variables are split irreducibly as
\begin{eqnarray}
\dot{u}^{a} &=& \mathcal{A} e^{a} + \mathcal{A}^{a}, \\
\omega^{a} &=& \Omega e^{a} + \Omega^{a}, \\
\sigma_{ab} &=& \Sigma\left(e_{a} e_{b} - \frac{1}{2} N_{ab}\right) + 2\Sigma_{(a} e_{b)} + \Sigma_{ab}, 
\end{eqnarray}
respectively, using (\ref{decomp1}) and (\ref{decomp2}). 
The covariant derivative of ${e^{a}}$ is given by
\begin{eqnarray}
D_{a} e_{b} &=& e_{a} a_{b} + \frac{1}{2}\phi N_{ab} + \xi\varepsilon_{ab} + \zeta_{ab}.
\end{eqnarray}
where traveling along ${e^{a}}$, ${a_{a}}$ is the sheet acceleration, ${\phi}$ is the sheet expansion, ${\xi}$ is the vorticity of ${e^{a}}$ (the twisting of the sheet) and ${\zeta_{ab}}$ is the shear of ${e^{a}}$ (the distortion of the sheet). The 1+1+2 split of the full covariant derivatives of ${u^{a}}$ and ${e^{a}}$ are as follows
\begin{eqnarray} \label{2.delAUb}
\nabla_{a} u_{b} &=& -u_{a}\left(\mathcal{A} e_{b} + \mathcal{A}_{b}\right) + e_{a} e_{b} \left(\frac{1}{3}\Theta + \Sigma \right) + e_{a}\left(\Sigma_{b} + \varepsilon_{bc}\Omega^{c}\right)  \nonumber\\
			& &+ \left(\Sigma_{a} - \varepsilon_{ac}\Omega^{c}\right) e_{b}+ N_{ab}\left(\frac{1}{3}\Theta - \frac{1}{2}\Sigma\right) + \Omega\varepsilon_{ab} + \Sigma_{ab}, \\
			\label{2.delAEb}
\nabla_{a} e_{b} &=& -\mathcal{A} u_{a} u_{b} - u_{a}\alpha_{b} + \left(\frac{1}{3}\Theta + \Sigma \right)e_{a} u_{b} + \left(\Sigma_{a} - \varepsilon_{ac}\Omega^{c}\right)u_{b}    \nonumber\\
			& & + e_{a} a_{b} +  \frac{1}{2}\phi N_{ab}+ \xi\varepsilon_{ab} + \zeta_{ab},			
\end{eqnarray}
where ${\mathcal{A}_{a} \equiv \dot{u}_{\bar{a}}, \,\alpha_{a} \equiv \dot{e}_{\bar{a}} \textrm{  and  } \varepsilon_{ab}}$ is the natural 2-volume element carried by the sheet. The bar on indices denotes projections on the sheet. A more detailed review of the formalism can be found in \cite{clarkson2007}.


\section{The 1+1+2 Kerr quantities and equations}
We apply the 1+1+2 formalism to the Kerr metric and find the following quantities confirmed by Maple. 
The spatial unit vector is defined as 
\begin{eqnarray}
e^{a} &=& \left[0, \sqrt{\frac{r^{2} - 2mr + a^{2}}{r^{2} + a^{2}\chi^{2}}}, 0, 0 \right]  \text{where } e^{a}e_{a} = 1. 
\end{eqnarray}

The set of Kerr 1+1+2 geometric variables is given by 
\begin{eqnarray}
&&\left\{ \mathcal{A}, \Omega, \mathcal{E}, \mathcal{H}, \phi, \xi, \mathcal{A}_{a}, \Omega_{a}, \alpha_{a},  a_{a}, \mathcal{E}_{a}, \mathcal{H}_{a}, \zeta_{ab}, \mathcal{E}_{ab}, \mathcal{H}_{ab}\right\},
\end{eqnarray}
and they have the following quantities 
 {\allowdisplaybreaks{\begin{eqnarray}
\mathcal{A}_{a} &=& N_{ab}\dot{u}^{b} \nonumber\\
			&=& \left[0, 0, \frac{2ma^{2}r\chi}{\mathcal{JK}}, 0\right], \\
\mathcal{A} &=& e^{a}\dot{u}_{a} \nonumber\\
 		   &=& \frac{-m\mathcal{D}\sqrt{\mathcal{L}}}{\mathcal{J}\sqrt{\mathcal{K}^{3}}},\\
\Omega &=& e^{a}\omega_{a} \nonumber\\
	      &=& \frac{-2mar\chi \sqrt{\mathcal{L}}}{\mathcal{J}\sqrt{\mathcal{K}^{3}}}, \\
\Omega_{a} &=& N_{ab}\omega^{b} \nonumber\\
		   &=& \left[0, 0, \frac{-ma\mathcal{D}}{\mathcal{JK}}, 0\right], \\
\varepsilon_{ab} &=& \varepsilon_{abc}e^{c} \nonumber\\
			 &=& \begin{bmatrix}
0 & 0 & 0 & 0\\
0 & 0 & 0 & 0\\
0 & 0 & 0 & -\mathcal{K}\sqrt{\frac{\mathcal{L}}{\mathcal{J}}}\\
0 & 0 & \mathcal{K}\sqrt{\frac{\mathcal{L}}{\mathcal{J}}}& 0
\end{bmatrix},\\
\mathcal{E} &=& E_{ab}e^{a}e^{b} \nonumber\\
		  &=& \frac{-mr\mathcal{P}\left(a^{2}\chi^{2} - 3a^{2} + 4mr - 2r^{2}\right)}{\mathcal{J}\mathcal{K}^{3}},\\
\mathcal{E}_{a} &=& N_{a}{}^{b}E_{bc}e^{c} \nonumber\\
			 &=& \left[0, 0, \frac{3ma^{2}\chi\mathcal{Q}\sqrt{\mathcal{L}}}{\mathcal{J}\sqrt{\mathcal{K}^{5}}}, 0\right], \\
\mathcal{E}_{ab} &=& E_{ab} - \mathcal{E}e_{a}e_{b} + \frac{1}{2}\mathcal{E}N_{ab} - \mathcal{E}_{a}e_{b} - \mathcal{E}_{b}e_{a} \nonumber\\
			   &=& \begin{bmatrix}
0 & 0 & 0 & 0\\
0 & 0 & 0 & 0\\
0 & 0 & -\frac{3}{2}\frac{ma^{2}r\mathcal{P}}{\mathcal{J}\mathcal{K}^{2}} & 0\\
0 & 0 & 0 & \frac{3}{2}\frac{ma^{2}r\mathcal{L}\mathcal{M}^{2}\mathcal{P}}{\mathcal{J}^2\mathcal{K}^{2}}
\end{bmatrix},\\  
\mathcal{H} &=& H_{ab}e^{a}e^{b} \nonumber\\
		   &=& \frac{-ma\chi\mathcal{Q}\left(a^{2}\chi^{2} - 3a^{2} + 4mr - 2r^{2}\right)}{\mathcal{J}\mathcal{K}^3},\\ 
\mathcal{H}_{a} &=& N_{a}{}^{b}H_{bc}e^{c} \nonumber\\
			 &=& \left[0, 0, -\frac{3mar\mathcal{P}\sqrt{\mathcal{L}}}{\mathcal{J}\sqrt{\mathcal{K}^{5}}}, 0\right], \\
\mathcal{H}_{ab} &=& H_{ab} - \mathcal{H}e_{a}e_{b} + \frac{1}{2}\mathcal{H}N_{ab} - \mathcal{H}_{a}e_{b} - \mathcal{H}_{b}e_{a} \nonumber \\
	  &=& \begin{bmatrix}
0 & 0 & 0 & 0\\
0 & 0 & 0 & 0\\
0 & 0 & -\frac{3}{2}\frac{ma^{3}\chi\mathcal{Q}}{\mathcal{J}\mathcal{K}^2} & 0\\
0 & 0 & 0 & \frac{3}{2}\frac{ma^{3}\chi\mathcal{L}\mathcal{M}^{2}\mathcal{Q}}{\mathcal{J}^{2}\mathcal{K}^{2}}
\end{bmatrix},\\
D_{b}e_{a} &=& h^{e}{}_{b} h^{c}{}_{a}\nabla_{e} e_{c}, \\
a_{a} &=& e^{c}D_{c} e_{a} \nonumber\\
	 &=& \left[0, 0, \frac{-a^{2}\chi}{\mathcal{K}}, 0\right], \\
\delta_{b} e_{a} &=& N_{b}{}^{c} N_{a}{}^{d}D_{c}e_{d}, \\
\phi &=& \delta_{a} e^{a} \nonumber\\
	&=& \frac{\mathcal{W} }{\mathcal{J}\sqrt{\mathcal{K}^3\mathcal{L}}},\\
	\xi &=& \frac{1}{2}\varepsilon^{ab}\delta_{a} e_{b} = 0, \\
\zeta_{ab} &=& \delta_{\{a} e_{b\}} \nonumber\\
		  &=& \begin{bmatrix}
0 & 0 & 0 & 0\\
0 & 0 & 0 & 0\\
0 & 0 & -\frac{1}{2}\frac{a^{2}\left(m-r\right)\sqrt{\mathcal{K}}}{\mathcal{J}\sqrt{\mathcal{L}}} & 0\\
0 & 0 & 0 & \frac{1}{2}\frac{a^{2}\left(m-r\right)\mathcal{M}^{2}\sqrt{\mathcal{LK}}}{\mathcal{J}^{2}}
\end{bmatrix},\\
\alpha_{a} &=& N_{ab}u^{c}\nabla_{c} e^{b} \nonumber\\
		&=& \left[0, 0, 0, \frac{ma\mathcal{MD}\sqrt{\mathcal{L}}}{\sqrt{\mathcal{J}^{3}}\mathcal{K}} \right]. 
\end{eqnarray}}}		
where 
\begin{eqnarray}	
\mathcal{W} &=& 2r^{3}\left(r-2m\right)^{2} + a^{4}\chi^{2}\left(m + r - m\chi^{2} + r\chi^{2}\right)  \nonumber\\
	& & + a^{2}r^{2}\left(-3m+r+\chi^{2}\left(3r-5m\right)\right). \n
\end{eqnarray}
All identities that are listed in Appendix B corresponding to the variables have been confirmed in Maple. The 1+1+2 decomposition method is an extension for the 1+3 decomposition method; hence the quantities ${\Theta}$ (expansion scalar) and ${\sigma}$-terms (shear) found to be zero in the previous section remain zero. We note that the sheet twist ${(\xi)}$ is found to be zero. The full set of the 1+1+2 Kerr equations for the above covariant variables are obtained by applying the 1+1+2 decomposition procedure to the 1+3 equations, and also by covariantly splitting the Ricci identities for ${e^{a}}$ as follows
\begin{equation} \label{rabc}
R_{abc} \equiv 2\nabla_{[a}\nabla_{b]}e_{c} - R_{abcd}e^{d} = 0,
\end{equation} 
where $R_{abcd}$ is the Riemann curvature tensor. Splitting \eqref{rabc} using ${u^{a}}$ and ${e^{a}}$ the evolution (along ${u^{a}}$) and propagation (along $e^{a}$) equations below are obtained. We present the full set of the 1+1+2 Kerr equations according to Clarkson \cite{clarkson2007}. 

\subsection{Evolution equations}
The evolution equations for ${\phi, \xi}$ and ${\zeta_{ab}}$ are obtained from the projection of $u^a R_{abc}$ as follows\\
${u^{a}N^{bc}R_{abc}}$:
\begin{equation} \label{faulty1}
\diamond\dot{\phi} = \delta_{a}\alpha{^a} + \alpha_{a}\left(\mathcal{A}^{a} - a^{a}\right) - \varepsilon_{ab}\Omega^{b}\left(a^{a} - \mathcal{A}^{a}\right).
\end{equation}
${u^{a}\varepsilon^{bc}R_{abc}}$:
\begin{equation}
\dot{\xi} = 0 = \left(\mathcal{A} - \frac{1}{2}\phi\right)\Omega + \frac{1}{2}\left(a^{a} + \mathcal{A}^{a}\right)\left[\Omega_{a} + \varepsilon_{ab}\alpha^{b}\right] + \frac{1}{2}\varepsilon_{ab}\delta^{a}\alpha^{b} + \frac{1}{2}\mathcal{H}.
\end{equation}
${u^{c}R_{c\{ab\}}}$:
\begin{eqnarray}
\dot{\zeta}_{\{ab\}} &=& \Omega\varepsilon_{c\{a}\zeta_{b\}}{}^{c} + \delta_{\{a}\alpha_{b\}} + \mathcal{A}_{\{a} \alpha_{b\}} - a_{\{a}\alpha_{b\}} + \mathcal{A}_{\{a}\varepsilon_{b\}d}\Omega^{d} \nonumber\\
& & + a_{\{a}\varepsilon_{b\}d}\Omega^{d} - \varepsilon_{c\{a}\mathcal{H}_{b\}}{}^{c}.
\end{eqnarray}
Not all information needed to determine the complete set of 1+1+2 equations is contained in $R_{abc}$. Hence the 1+1+2 decomposition of the standard 1+3 equations is used to obtain the remaining evolution equations given below. \\
Vorticity evolution equation:
\begin{equation}
\dot{\Omega} = \frac{1}{2}\varepsilon_{ab}\delta^{a}\mathcal{A}^{b} + \Omega_{a}\alpha^{a}.
\end{equation}
Shear evolution equation:
\begin{equation}
0 = \delta_{\{a}\mathcal{A}_{b\}} + \mathcal{A}_{\{a}\mathcal{A}_{b\}} - \Omega_{\{a}\Omega_{b\}} + \mathcal{A}\zeta_{ab} - \mathcal{E}_{ab}.
\end{equation}

\subsection{Mixture of propagation and evolution equations}
A mixture of propagation and evolution equations is obtained by either projecting ${R_{abc}}$ as indicated or by a further decomposition of the 1+3 equations.\\
${u^{a}e^{b}R_{ab\bar{c}} = e^{a}u^{b}R_{ab\bar{c}}}$:
\begin{eqnarray}
\hat{\alpha}_{\bar{a}} - \dot{a}_{\bar{a}} &=& -\left(\frac{1}{2}\phi + \mathcal{A}\right)\alpha_{a} + \varepsilon_{ab}\Omega^{b}\left(\frac{1}{2}\phi - \mathcal{A}\right) \nonumber\\
& & + \zeta_{ab}\left(\varepsilon^{bc}\Omega_{c} - \alpha^{b}\right) - \varepsilon_{ab}\mathcal{H}^{b}.
\end{eqnarray}
Raychaudhuri equation:
\begin{eqnarray}
\hat{\mathcal{A}} &=& -\delta_{a}\mathcal{A}^{a} - \mathcal{A}\left(\mathcal{A} + \phi\right) + \mathcal{A}^{a}\left(a_{a} - \mathcal{A}_{a}\right) - 2\Omega^{2} -2\Omega_{a}\Omega^{a}. 
\end{eqnarray}
Vorticity evolution equation:
\begin{equation}
\dot{\Omega}_{\bar{a}} + \frac{1}{2}\varepsilon_{ab}\hat{\mathcal{A}}^{b} = -\Omega\alpha_{a} + \frac{1}{2}\varepsilon_{ab}\left(-\mathcal{A}a^{b} + \delta^{b}\mathcal{A} - \frac{1}{2}\phi\mathcal{A}^{b}\right) - \frac{1}{2}\varepsilon_{ab}\zeta^{bc}\mathcal{A}_{c}.
\end{equation}
Shear evolution equation:
\begin{eqnarray}
\label{faulty3}
\diamond -\frac{2}{3}\hat{\mathcal{A}} &=& \frac{2}{3}\mathcal{A}^{2} - \frac{1}{3}\phi\mathcal{A} - \frac{2}{3}\Omega^{2} - \frac{1}{3}\delta_{a}\mathcal{A}^{a} - \frac{2}{3}\mathcal{A}_{a}a^{a} - \frac{1}{3}\mathcal{A}_{a}\mathcal{A}^{a} \nonumber\\
& & + \frac{1}{3}\Omega_{a}\Omega^{a} - \mathcal{E}, \\
-\frac{1}{2}\hat{\mathcal{A}}_{\bar{a}} &=& \frac{1}{2}\delta_{a}\mathcal{A} + \mathcal{A}_{a}\left(\mathcal{A} - \frac{1}{4}\phi\right) + \frac{1}{2}\mathcal{A}a_{a} - \Omega\Omega_{a} -\frac{1}{2}\zeta_{ab}\mathcal{A}^{b} - \mathcal{E}_{a}.
\end{eqnarray}

Additionally the magnetic and electric Weyl evolution equations are listed below. \\

\noindent Electric Weyl evolution equations:
\begin{eqnarray}
\dot{\mathcal{E}} &=& \varepsilon_{ab}\delta^{a}\mathcal{H}^{b} + \mathcal{E}^{a}\left(2\alpha_{a} - \varepsilon_{ab}\Omega^{b}\right) + 2\varepsilon_{ab}\mathcal{A}^{a}\mathcal{H}^{b} + \varepsilon_{ab}\mathcal{H}^{bc}\zeta^{a}{}_{c}, \\
\label{faulty4}
\diamond\dot{\mathcal{E}}_{\bar{a}} + \frac{1}{2}\varepsilon_{ab}\hat{\mathcal{H}}^{b} &=& \frac{3}{4}\varepsilon_{ab}\delta^{b}\mathcal{H} + \frac{1}{2}\varepsilon_{bc}\delta^{b}\mathcal{H}^{c}{}_{a} + \frac{3}{4}\mathcal{E}\varepsilon_{ab}\Omega^{b} + \frac{3}{2}\mathcal{H}\varepsilon_{ab}\mathcal{A}^{b} - \frac{3}{2}\mathcal{E}\alpha_{a}  \nonumber \\
& & - \frac{3}{4}\mathcal{H}\varepsilon_{ab}a^{b} - \frac{1}{2}\Omega\varepsilon_{ab}\mathcal{E}^{b} - \frac{1}{4}\phi\varepsilon_{ab}\mathcal{H}^{b} - \mathcal{A}\varepsilon_{ab}\mathcal{H}^{b} + \mathcal{E}_{ab}\alpha^{b} \nonumber\\
& &  - \frac{1}{2}\mathcal{E}_{ab}\varepsilon^{bc}\Omega_{c} + \frac{3}{2}\zeta_{ab}\varepsilon^{bc}\mathcal{H}_{c} - \mathcal{H}_{ab}\varepsilon^{bc}\mathcal{A}_{c} +  \frac{1}{2}\varepsilon_{ab}a^{c}\mathcal{H}^{b}{}_{c} \nonumber\\
& &+ \varepsilon_{ab}e_{c}\delta^{b}\mathcal{H}^{c}, \\
\dot{\mathcal{E}}_{\{ab\}} - \varepsilon_{c\{a}\hat{\mathcal{H}}_{b\}}{}^{c} &=& -\varepsilon_{c\{a}\delta^{c}\mathcal{H}_{b\}} - \frac{3}{2}\mathcal{H}\varepsilon_{c\{a}\zeta_{b\}}{}^{c} + \Omega\varepsilon_{c\{a}\mathcal{E}_{b\}}{}^{c} \nonumber\\
&& + \varepsilon_{c\{a}\mathcal{H}_{b\}}{}^{c}\left(\frac{1}{2}\phi + 2\mathcal{A}\right) - 2\alpha_{\{a}\mathcal{E}_{b\}} - \varepsilon_{c\{a}\mathcal{E}_{b\}}\Omega^{c} \nonumber\\ 
&&+ 2\varepsilon_{c\{a}\mathcal{H}_{b\}}\left(a^{c} - \mathcal{A}^{c}\right) + \varepsilon_{c\{a}\mathcal{H}_{b\}d}\zeta^{cd}.
\end{eqnarray}
\noindent Magnetic Weyl evolution equations:
\begin{equation}
\dot{\mathcal{H}} = -\varepsilon_{ab}\delta^{a}\mathcal{E}^{b} - 2\varepsilon_{ab}\mathcal{A}^{a}\mathcal{E}^{b} + \mathcal{H}^{a}\left(2\alpha_{a} - \varepsilon_{ab}\Omega^{b}\right) - \frac{1}{2}\varepsilon_{ab}\mathcal{E}^{bc}\zeta^{a}{}_{c},
\end{equation}
\begin{eqnarray}
\label{faulty5}
\diamond \dot{\mathcal{H}}_{\bar{a}} - \frac{1}{2}\varepsilon_{ab}\hat{\mathcal{E}}^{b} &=& -\frac{3}{4}\varepsilon_{ab}\delta^{b}\mathcal{E} - \frac{1}{2}\varepsilon_{bc}\delta^{b}\mathcal{E}^{c}{}_{a} + \frac{3}{4}\mathcal{H}\varepsilon_{ab}\Omega^{b} - \frac{3}{2}\mathcal{E}\varepsilon_{ab}\mathcal{A}^{b}  \nonumber \\
& &  - \frac{3}{2}\mathcal{H}\alpha_{a}  + \frac{3}{4}\mathcal{E}\varepsilon_{ab}a^{b} + \frac{1}{4}\phi\varepsilon_{ab}\mathcal{E}^{b} + \mathcal{A}\varepsilon_{ab}\mathcal{E}^{b} - \frac{1}{2}\Omega\varepsilon_{ab}\mathcal{H}^{b} \nonumber \\
& &+ \frac{3}{2}\varepsilon_{ab}\zeta^{bc}\mathcal{E}_{c}  + \mathcal{E}_{ab}\varepsilon^{bc}\mathcal{A}_{c} + \mathcal{H}_{ab}\alpha^{b} - \frac{1}{2}\mathcal{H}_{ab}\varepsilon^{bc}\Omega_{c} \nonumber\\
& & - \frac{1}{2}\varepsilon_{ab}a^{c}\mathcal{E}^{b}{}_{c}  - \varepsilon_{ab}e_{c}\delta^{b}\mathcal{E}^{c}, \\
\label{faulty6}
\diamond \dot{\mathcal{H}}_{\{ab\}} + \varepsilon_{c\{a}\hat{\mathcal{E}}_{b\}}{}^{c} &=& \varepsilon_{c\{a}\delta^{c}\mathcal{E}_{b\}} + \frac{3}{2}\mathcal{E}\varepsilon_{c\{a}\zeta_{b\}}{}^{c} - \frac{1}{2}\phi\varepsilon_{c\{a}\mathcal{E}_{b\}}{}^{c} - 2\mathcal{A}\varepsilon_{c\{a}\mathcal{E}_{b\}}{}^{c} \nonumber\\
& & + \Omega\varepsilon_{c\{a}\mathcal{H}_{b\}}{}^{c} -\varepsilon_{c\{a}\mathcal{H}_{b\}}\Omega^{c} + 2\mathcal{E}_{\{a}\varepsilon_{b\}c}a^{c} - 2\mathcal{E}_{\{a}\varepsilon_{b\}c}\mathcal{A}^{c} \nonumber\\
& & - 2\alpha_{\{a}\mathcal{H}_{b\}} - \varepsilon_{c\{a}\mathcal{E}_{b\}d}\zeta^{cd}.
\end{eqnarray}

\subsection{Propagation equations}
Following a similar procedure, the propagation and constraint equations are obtained by either projecting ${R_{abc}}$ as indicated or from projections of the 1+3 constraint equations in section 4. \\
${e^{a}N^{bc}R_{abc}}$:
\begin{eqnarray}
\hat{\phi} &=& -\frac{1}{2}\phi^{2} + \delta_{a}a^{a} - a_{a}a^{a} - \zeta_{ab}\zeta^{ab} + 2\varepsilon_{ab}\alpha^{a}\Omega^{b} + \Omega_{a}\Omega^{a} - \mathcal{E}.
\end{eqnarray}
${e^{a}\varepsilon^{bc}R_{abc}}$:
\begin{eqnarray} \label{faulty7}
\diamond\hat{\xi} = 0 = \frac{1}{2}\varepsilon_{ab}\delta^{a}a^{b} + \alpha_{a}\Omega^{a}.
\end{eqnarray}
${e^{a}R_{a\{bc\}}}$:
\begin{eqnarray}
\hat{\zeta}_{\{ab\}} &=& -\phi\zeta_{ab} - \zeta^{c}{}_{\{a}\zeta_{b\}c} + \delta_{\{a}a_{b\}} - a_{\{a}a_{b\}} + 2\alpha_{\{a}\varepsilon_{b\}c}\Omega^{c} - \Omega_{\{a}\Omega_{b\}} - \mathcal{E}_{ab}.
\end{eqnarray}
Additionally, the divergence equations for the shear, vorticity and the electric and magnetic Weyl parts are written below. \\
Shear divergence equations:
\begin{eqnarray}
0 &=& -\varepsilon_{ab}\delta^{a}\Omega^{b} - 2\varepsilon_{ab}\mathcal{A}^{a}\Omega^{b}, \\
-\varepsilon_{ab}\hat{\Omega}^{b} &=& -\varepsilon_{ab}\delta^{b}\Omega + \varepsilon_{ab}\Omega^{b}\left(\frac{1}{2}\phi + 2\mathcal{A}\right) + \Omega\varepsilon_{ab}\left(a^{b} - 2\mathcal{A}^{b}\right) + \varepsilon_{ab}\zeta^{bc}\Omega_{c}.
\end{eqnarray}
Vorticity divergence equations:
\begin{eqnarray}
\hat{\Omega} &=& -\delta_{a}\Omega^{a} + \Omega\left(\mathcal{A} - \phi\right) + \Omega^{a}\left(a_{a} + \mathcal{A}_{a}\right), \\
0 &=& -\varepsilon_{c\{a}\delta^{c}\Omega_{b\}} - \Omega\varepsilon_{c\{a}\zeta_{b\}}{}^{c} - 2\varepsilon_{c\{a}\Omega_{b\}}\mathcal{A}^{c} - \varepsilon_{c\{a}\mathcal{H}_{b\}}{}^{c}.
\end{eqnarray}
Electrical Weyl divergence equations:
\begin{eqnarray}
\hat{\mathcal{E}} &=& -\delta_{a}\mathcal{E}^{a} - \frac{3}{2}\phi\mathcal{E} + 3\Omega\mathcal{H} + 2\mathcal{E}_{a}a^{a} + 3\Omega_{a}\mathcal{H}^{a} + \mathcal{E}_{ab}\zeta^{ab}, \\
\hat{\mathcal{E}}_{\bar{a}} &=& \frac{1}{2}\delta_{a}\mathcal{E} - \delta^{b}\mathcal{E}_{ab} - \frac{3}{2}\mathcal{H}\Omega_{a} - \frac{3}{2}\mathcal{E}a_{a} - \frac{3}{2}\phi\mathcal{E}_{a} + 3\Omega\mathcal{H}_{a} - \zeta_{ab}\mathcal{E}^{b} \nonumber\\
& & + \mathcal{E}_{ab}a^{b} + 3\mathcal{H}_{ab}\Omega^{b}.
\end{eqnarray}
Magnetic Weyl divergence equations:
\begin{eqnarray}
\hat{\mathcal{H}} &=& -\delta_{a}\mathcal{H}^{a} - \frac{3}{2}\phi\mathcal{H} - 3\Omega\mathcal{E} + 2\mathcal{H}_{a}a^{a} -3\Omega_{a}\mathcal{E}^{a} + \zeta_{ab}\mathcal{H}^{ab}, \\
\hat{\mathcal{H}}_{\bar{a}} &=& \frac{1}{2}\delta_{a}\mathcal{H} - \delta^{b}\mathcal{H}_{ab} + \frac{3}{2}\mathcal{E}\Omega_{a} - \frac{3}{2}\mathcal{H}a_{a} - 3\Omega\mathcal{E}_{a} - \frac{3}{2}\phi\mathcal{H}_{a} + \mathcal{H}_{ab}a^{b} \nonumber\\
& &- \zeta_{ab}\mathcal{H}^{b} - 3\mathcal{E}_{ab}\Omega^{b}.
\end{eqnarray}

\subsection{Constraint equations}
${\varepsilon^{ab}u^{c}R_{abc}}$:
\begin{eqnarray}
\delta_{a}\Omega^{a} &=& \Omega\left(2\mathcal{A} - \phi\right) + \mathcal{H}.
\end{eqnarray}
${N^{bc}R_{\bar{a}bc}}$:
\begin{eqnarray}
\frac{1}{2}\delta_{a}\phi - \delta^{b}\zeta_{ab} &=& -\Omega\Omega_{a} + 2\Omega\varepsilon_{ab}\alpha^{b} - \mathcal{E}_{a}.
\end{eqnarray}
${e^{a}u^{c}R_{a\bar{b}c}}$:
\begin{eqnarray}
2\varepsilon_{ab}\delta^{b}\Omega &=& \phi\varepsilon_{ab}\Omega^{b} - 4\Omega\varepsilon_{ab}\mathcal{A}^{b} + 2\varepsilon_{ab}\zeta^{bc}\Omega_{c} - 2\varepsilon_{ab}\mathcal{H}^{b}.
\end{eqnarray}

All the above equations have been validated in Maple. We note that we reworked some of the equations in \cite{clarkson2007} and these are denoted by ${\diamond}$. A comparison between Clarkson's result and the corrected result can be found in Appendix C. We emphasise that the linear perturbation results in \cite{clarksonbarrett} and \cite{clarkson2007} still hold as the identified errors are higher order terms.


\section{Conclusion}

In this paper we have given a complete 1+1+2 semi-tetrad covariant description of the Kerr spacetime outside the ergoregion. Since there are no matter quantities we explicitly wrote down all the geometrical quantities and the evolution and propagation equations they satisfy. During this process we identified some errors in the equations of the previous 1+1+2 decomposition performed in \cite{clarksonbarrett} and \cite{clarkson2007} which we corrected in this paper specifically for the Kerr spacetime. However the affected terms in those equations were higher order terms. Hence the linear perturbation results in \cite{clarksonbarrett} and \cite{clarkson2007} remain unaffected. 

As far as we are aware, the 1+1+2 description of Kerr geometry presented in this paper is the first comprehensive treatment. Furthermore this highlights the role of geometrical variables beautifully. As a natural generalisation of the linear perturbation of Schwarzschild geometry, where the Regge-Wheeler tensor was found \cite{clarksonbarrett}, we can perform a similar but extensive calculation to find the corresponding Teukolsky tensor for rotating geometry. 

In the future we can use this detailed geometrical description of the Kerr spacetime to find different physical properties of the Kerr spacetime that would have been more difficult to do using the usual coordinate approach or a standard 1+3 decomposition which was also performed in this paper. 


\appendix
\begin{appendix}
	\section{Additional 1+3 definitions}
We write down important identities and definitions in the 1+3 formalism. Note that
\begin{eqnarray}
\varepsilon_{abc} &=& \sqrt{|\textrm{det } g|}\sss\delta^0{}_{[a}\sss \delta^1{}_{b}\sss\delta^2{}_{c}\sss\delta^3{}_{d]} \sss u^{d}, \nonumber\\
\varepsilon_{abc}\varepsilon^{def} &=& 3!  h^d{}_{\left[ a \right.}  h^e{}_{b}  h^f{}_{\left. c \right]}, \nonumber\\
\varepsilon_{abc}\varepsilon^{dec} &=& 2 h^d{}_{\left[ a \right.} h^e{}_{\left. b\right]}. \nonumber
\end{eqnarray}
Angle brackets denote orthogonal projections of covariant time derivatives along ${u^{a}}$ as well as represent the projected, symmetric and trace-free part of tensors as follows
\begin{eqnarray}
v_{<a>} = h^{b}{}_{a}\dot{V}_{b}, \quad\quad Z_{<ab>} = \left(h^{c}{}_{(a} h^{d}{}_{b)} - \frac{1}{3}h_{ab} h^{cd}\right) Z_{cd}.
\end{eqnarray}


\section{Additional 1+1+2 definitions}
We write down the definitions of important components in the 1+1+2 formalism.
Any spacetime 3-vector ${\Phi^{a}}$ can be irreducibly split into ${\chi}$, a scalar component along ${e^{a}}$, and a 2-vector ${\chi^{a}}$, which is a sheet component orthogonal to ${e^{a}}$, as follows
\begin{eqnarray} \label{decomp1}
\Phi^{a} = \chi e^{a} + \chi^{a} \quad \textrm{where} \quad  \chi \equiv \Phi_{a} e^{a} \quad  \textrm{and} \quad \chi^{a} \equiv N^{ab} \Phi_{b} \equiv \Phi^{\bar{a}},
\end{eqnarray}
where the bar on a particular index denotes projection with ${N_{ab}}$ on that index such that the vector or tensor lies on the sheet. Similarly we can split a projected, symmetric, trace-free tensor ${\Phi_{ab}}$ into scalar, 2-vector and 2-tensor parts as follows
\begin{equation} \label{decomp2}
\Phi_{ab} = \Phi_{<ab>} = \chi\left(e_{a} e_{b} - \frac{1}{2} N_{ab}\right) + 2\chi_{(a} e_{b)} + \chi_{ab},
\end{equation}
where the components
\begin{eqnarray}
\chi &\equiv& e^{a} e^{b} \Phi_{ab} = -N^{ab}\Phi_{ab}, \nonumber\\
\chi_{a} &\equiv& N_{a}{}^{b} e^{c}\Phi_{bc}, \nonumber\\
\chi_{ab} &\equiv& \chi_{\{ab\}} = \left(N_{(a}{}^{c} N_{b)}{}^{d} - \frac{1}{2} N_{ab} N^{cd}\right)\Phi_{cd},
\end{eqnarray}
are defined. The curly brackets denote the part of the tensor that is projected, symmetric and trace-free, with respect to ${e^{a}}$.  

We mention that in general the dot, hat and delta derivatives do not commute. The commutation relations for any scalar ${\kappa}$ are 
\begin{eqnarray}
\label{2.firstComScalar}
\hat{\dot{\kappa}} - \dot{\hat{\kappa}} &=& -\mathcal{A}\dot{\kappa} + \left(\frac{1}{3}\Theta + \Sigma\right)\hat{\kappa}  + \left(\Sigma_{a} + \varepsilon_{ab}\Omega^{b} - \alpha_{a} \right)\delta^{a}\kappa, \\
\delta_{a}\dot{\kappa} - \left(\delta_{a}\kappa\right)^{\cdot}_{\perp} &=& -\mathcal{A}_{a}\dot{\kappa} + \left(\alpha_{a} + \Sigma_{a} - \varepsilon_{ab}\Omega^{b} \right)\hat{\kappa}  \nonumber\\
						& & + \left(\frac{1}{3}\Theta - \frac{1}{2}\Sigma\right)\delta_{a}\kappa + \left(\Sigma_{ab} + \Omega\varepsilon_{ab}\right)\delta^{b}\kappa, \\
\delta_{a}\hat{\kappa} - \left(\widehat{\delta_{a}\kappa}\right)_{\perp} &=& -2\varepsilon_{ab}\Omega^{b}\dot{\kappa} + a_{a}\hat{\kappa} + \frac{1}{2}\phi\delta_{a}\kappa + \left(\zeta_{ab} + \xi\varepsilon_{ab}\right)\delta^{b}\kappa,\\
\label{2.lastComScalar}
\delta_{[a}\delta_{b]}\kappa &=& \varepsilon_{ab}\left(\Omega\dot{\kappa} - \alpha\hat{\kappa}\right),
\end{eqnarray}
where ${\perp}$ denotes projection onto the sheet. 

\section{Corrections to Clarkson equations}

This appendix contains the equations of Clarkson \cite{clarkson2007} that contained errors and their subsequent corrections in the Kerr formalism. We note once again that the errors were detected in higher order terms hence the results presented in \cite{clarksonbarrett} and \cite{clarkson2007} remained unaffected. These equations have been checked using the mathematical software Maple and GRTensor. The corrected terms have been enclosed in boxes below.
\begin{enumerate}
\item Evolution equation (\ref{faulty1})
\\
${u^{a}N^{bc}R_{abc}}$: 
\\
Clarkson:
\begin{eqnarray} 
\dot{\phi} &=& \delta_{a}\alpha{^a} + \mathcal{A}^{a}\left(\alpha_{a} - a_{a}\right) - \varepsilon_{ab}\Omega^{b}\left(a^{a} - \mathcal{A}^{a}\right).
\end{eqnarray} 
Correction:
\begin{eqnarray} \dot{\phi} &=& \delta_{a}\alpha{^a} + \boxed{\alpha_{a}\left(\mathcal{A}^{a} - a^{a}\right)} - \varepsilon_{ab}\Omega^{b}\left(a^{a} - \mathcal{A}^{a}\right).
\end{eqnarray}

\item Mixed equation (\ref{faulty3})
\\
Shear evolution: 
\\
Clarkson:
\begin{eqnarray}
-\frac{2}{3}\hat{\mathcal{A}} &=& \frac{2}{3}\mathcal{A}^{2} - \frac{1}{3}\phi\mathcal{A} - \frac{2}{3}\Omega^{2} - \frac{1}{3}\delta_{a}\mathcal{A}^{a} - \frac{2}{3}\mathcal{A}_{a}a^{a}  + \frac{1}{3}\mathcal{A}_{a}\mathcal{A}^{a} + \frac{1}{3}\Omega_{a}\Omega^{a} - \mathcal{E}. 
\end{eqnarray}
Correction:
\begin{eqnarray}
-\frac{2}{3}\hat{\mathcal{A}} &=& \frac{2}{3}\mathcal{A}^{2} - \frac{1}{3}\phi\mathcal{A} - \frac{2}{3}\Omega^{2} - \frac{1}{3}\delta_{a}\mathcal{A}^{a} - \frac{2}{3}\mathcal{A}_{a}a^{a} \boxed{- \frac{1}{3}\mathcal{A}_{a}\mathcal{A}^{a}} + \frac{1}{3}\Omega_{a}\Omega^{a} - \mathcal{E}.
\end{eqnarray}

\item Mixed equation (\ref{faulty4})
\\
Electric Weyl evolution: 
\\
Clarkson:
\begin{eqnarray}
\dot{\mathcal{E}}_{\bar{a}} + \frac{1}{2}\varepsilon_{ab}\hat{\mathcal{H}}^{b} &=& \frac{3}{4}\varepsilon_{ab}\delta^{b}\mathcal{H} + \frac{1}{2}\varepsilon_{bc}\delta^{b}\mathcal{H}^{c}{}_{a} + \frac{3}{4}\mathcal{E}\varepsilon_{ab}\Omega^{b} + \frac{3}{2}\mathcal{H}\varepsilon_{ab}\mathcal{A}^{b} - \frac{3}{2}\mathcal{E}\alpha_{a} - \frac{3}{4}\mathcal{H}\varepsilon_{ab}a^{b} \nonumber \\
& & - \frac{1}{2}\Omega\varepsilon_{ab}\mathcal{E}^{b} - \frac{1}{4}\phi\varepsilon_{ab}\mathcal{H}^{b} - \mathcal{A}\varepsilon_{ab}\mathcal{H}^{b} - \mathcal{E}_{ab}\alpha^{b} - \frac{1}{2}\mathcal{E}_{ab}\varepsilon^{bc}\Omega_{c} +  \frac{1}{2}\zeta_{ab}\varepsilon^{bc}\mathcal{H}_{c} \nonumber\\
& &- \mathcal{H}_{ab}\varepsilon^{bc}\mathcal{A}_{c}.
\end{eqnarray}
Correction:
\begin{eqnarray}
\dot{\mathcal{E}}_{\bar{a}} + \frac{1}{2}\varepsilon_{ab}\hat{\mathcal{H}}^{b} &=& \frac{3}{4}\varepsilon_{ab}\delta^{b}\mathcal{H} + \frac{1}{2}\varepsilon_{bc}\delta^{b}\mathcal{H}^{c}{}_{a} + \frac{3}{4}\mathcal{E}\varepsilon_{ab}\Omega^{b} + \frac{3}{2}\mathcal{H}\varepsilon_{ab}\mathcal{A}^{b} - \frac{3}{2}\mathcal{E}\alpha_{a} - \frac{3}{4}\mathcal{H}\varepsilon_{ab}a^{b} \nonumber \\
& & - \frac{1}{2}\Omega\varepsilon_{ab}\mathcal{E}^{b} - \frac{1}{4}\phi\varepsilon_{ab}\mathcal{H}^{b} - \mathcal{A}\varepsilon_{ab}\mathcal{H}^{b} \boxed{+ \mathcal{E}_{ab}\alpha^{b}} - \frac{1}{2}\mathcal{E}_{ab}\varepsilon^{bc}\Omega_{c} + \boxed{\frac{3}{2}\zeta_{ab}\varepsilon^{bc}\mathcal{H}_{c}} \nonumber\\
& &- \mathcal{H}_{ab}\varepsilon^{bc}\mathcal{A}_{c} \boxed{ +  \frac{1}{2}\varepsilon_{ab}a^{c}\mathcal{H}^{b}{}_{c} + \varepsilon_{ab}e_{c}\delta^{b}\mathcal{H}^{c}}.
\end{eqnarray}

\item Mixed equation (\ref{faulty5})
\\
Magnetic Weyl evolution: 
\\Clarkson:
\begin{eqnarray}
\dot{\mathcal{H}}_{\bar{a}} - \frac{1}{2}\varepsilon_{ab}\hat{\mathcal{E}}^{b} &=& -\frac{3}{4}\varepsilon_{ab}\delta^{b}\mathcal{E} - \frac{1}{2}\varepsilon_{bc}\delta^{b}\mathcal{E}^{c}{}_{a} + \frac{3}{4}\mathcal{H}\varepsilon_{ab}\Omega^{b} - \frac{3}{2}\mathcal{E}\varepsilon_{ab}\mathcal{A}^{b} - \frac{3}{2}\mathcal{H}\alpha_{a} \nonumber \\
& &+ \frac{3}{4}\mathcal{E}\varepsilon_{ab}a^{b} + \frac{1}{4}\phi\varepsilon_{ab}\mathcal{E}^{b} + \mathcal{A}\varepsilon_{ab}\mathcal{E}^{b} - \frac{1}{2}\Omega\varepsilon_{ab}\mathcal{H}^{b} + \frac{3}{2}\varepsilon_{ab}\zeta^{bc}\mathcal{E}_{c} \nonumber \\
& &+ \varepsilon_{ab}\mathcal{A}_{c}\zeta^{bc} + \mathcal{H}_{ab}\alpha^{b} - \frac{1}{2}\mathcal{H}_{ab}\varepsilon^{bc}\Omega_{c}.
\end{eqnarray}
Correction:
\begin{eqnarray}
\dot{\mathcal{H}}_{\bar{a}} - \frac{1}{2}\varepsilon_{ab}\hat{\mathcal{E}}^{b} &=& -\frac{3}{4}\varepsilon_{ab}\delta^{b}\mathcal{E} - \frac{1}{2}\varepsilon_{bc}\delta^{b}\mathcal{E}^{c}{}_{a} + \frac{3}{4}\mathcal{H}\varepsilon_{ab}\Omega^{b} - \frac{3}{2}\mathcal{E}\varepsilon_{ab}\mathcal{A}^{b} - \frac{3}{2}\mathcal{H}\alpha_{a} \nonumber \\
& &+ \frac{3}{4}\mathcal{E}\varepsilon_{ab}a^{b} + \frac{1}{4}\phi\varepsilon_{ab}\mathcal{E}^{b} + \mathcal{A}\varepsilon_{ab}\mathcal{E}^{b} - \frac{1}{2}\Omega\varepsilon_{ab}\mathcal{H}^{b} + \frac{3}{2}\varepsilon_{ab}\zeta^{bc}\mathcal{E}_{c} \nonumber \\
& &+ \boxed{\mathcal{E}_{ab}\varepsilon^{bc}\mathcal{A}_{c}} + \mathcal{H}_{ab}\alpha^{b} - \frac{1}{2}\mathcal{H}_{ab}\varepsilon^{bc}\Omega_{c} \boxed{- \frac{1}{2}\varepsilon_{ab}a^{c}\mathcal{E}^{b}{}_{c} - \varepsilon_{ab}e_{c}\delta^{b}\mathcal{E}^{c}}.
\end{eqnarray}

\item Mixed equation (\ref{faulty6})
\\
Magnetic Weyl evolution: 
\\
Clarkson:
\begin{eqnarray}
\dot{\mathcal{H}}_{\{ab\}} + \varepsilon_{c\{a}\hat{\mathcal{E}}_{b\}}{}^{c} &=& \varepsilon_{c\{a}\delta^{c}\mathcal{E}_{b\}} + \frac{3}{2}\mathcal{E}\varepsilon_{c\{a}\zeta_{b\}}{}^{c} - \frac{1}{2}\phi\varepsilon_{c\{a}\mathcal{E}_{b\}}{}^{c} - 2\mathcal{A}\varepsilon_{c\{a}\mathcal{E}_{b\}}{}^{c} - \Omega\varepsilon_{c\{a}\mathcal{H}_{b\}}{}^{c} \nonumber\\
& &- \Omega_{\{a}\varepsilon_{b\}c}\mathcal{H}^{c} - 2\alpha_{\{a}\mathcal{H}_{b\}} + 2\mathcal{E}_{\{a}\varepsilon_{b\}c}a^{c} + 2\mathcal{E}_{\{a}\varepsilon_{b\}c}\mathcal{A}^{c} \nonumber\\
&&  - \varepsilon_{c\{a}\mathcal{E}_{b\}d}\zeta^{cd}.
\end{eqnarray}
Correction:
\begin{eqnarray}
\dot{\mathcal{H}}_{\{ab\}} + \varepsilon_{c\{a}\hat{\mathcal{E}}_{b\}}{}^{c} &=& \varepsilon_{c\{a}\delta^{c}\mathcal{E}_{b\}} + \frac{3}{2}\mathcal{E}\varepsilon_{c\{a}\zeta_{b\}}{}^{c} - \frac{1}{2}\phi\varepsilon_{c\{a}\mathcal{E}_{b\}}{}^{c} - 2\mathcal{A}\varepsilon_{c\{a}\mathcal{E}_{b\}}{}^{c} \boxed{+ \Omega\varepsilon_{c\{a}\mathcal{H}_{b\}}{}^{c}} \nonumber\\
& & -\boxed{\varepsilon_{c\{a}\mathcal{H}_{b\}}\Omega^{c}} - 2\alpha_{\{a}\mathcal{H}_{b\}} + 2\mathcal{E}_{\{a}\varepsilon_{b\}c}a^{c} \boxed{- 2\mathcal{E}_{\{a}\varepsilon_{b\}c}\mathcal{A}^{c}} \nonumber\\ &&- \varepsilon_{c\{a}\mathcal{E}_{b\}d}\zeta^{cd}.
\end{eqnarray}

\item Propagation equation (\ref{faulty7})
\\
${e^{a}}\varepsilon^{bc}R_{abc}$: 
\\
Clarkson:
\begin{eqnarray}
\hat{\xi} = 0 = \frac{1}{2}\varepsilon_{ab}\delta^{a}a^{b} + \alpha_{a}\Omega^{a} + \frac{1}{2}a_{a}\Omega^{a}.
\end{eqnarray}
Correction:
\begin{eqnarray}
\hat{\xi} = 0 = \frac{1}{2}\varepsilon_{ab}\delta^{a}a^{b} + \alpha_{a}\Omega^{a}.
\end{eqnarray}

\end{enumerate}
					
\end{appendix}

\begin{acknowledgements} 
CH is supported by the DST-NRF Centre of Excellence in Mathematical and Statistical Sciences (CoE-MaSS) and the University of KwaZulu-Natal. Opinions expressed and conclusions arrived at are those of the author and are not necessarily to be attributed to the CoE-MaSS. RG is supported by the National Research Foundation (NRF), South Africa, and the University of KwaZulu-Natal. SDM acknowledges that this work was based on research supported by the South African Research Chair Initiative of the Department of Science and Technology and the National Research Foundation. 
\end{acknowledgements}

%

\end{document}